\documentstyle[12pt, iopfts]{ioplppt}

\begin{document}

\sloppy

\jl{2}

\title{Electron detachment from negative ions in bichromatic
laser field}

\author{M.~Yu.~Kuchiev\footnote{E-mail: kuchiev@newt.phys.unsw.edu.au}
 and V.~N.~Ostrovsky\footnote{
Permanent address: 
Institute of Physics, The University of St Petersburg,
198904 St.Petersburg, Russia; E-mail: Valentin.Ostrovsky@pobox.spbu.ru
}
}

\address{School of Physics, University of New South Wales, 
Sydney 2052, Australia}

%\maketitle

%**************************************************************************
\begin{abstract}

Negative ion detachment in two-colour laser field
is considered within the recent modification of Keldysh model
which makes it quantitatively reliable.
The general approach is illustrated by calculation of
angular differential detachment rates, partial rates
for particular ATD (Above Threshold Detachment) channels 
and total detachment rates for H$^-$ ion in bichromatic field 
with 1:2 frequency ratio. Both perturbative and strong field
regimes are examined. Polar asymmetry and phase effects are
quantitatively characterized with some new features revealed. 
Phase effects are found to result in 
a huge anisotropy factor $\sim 10^3$ in the electron
angular distribution in the perturbative regime.
%**************************************************
\end{abstract}

\section{Introduction}

Photoionization of atoms in bichromatic laser field had received
recently a considerable attention both in theory
(see, for instance, Baranova {\it et al}\/ 1990, 1992, 1993, 
Baranova and Zel'dovich 1991,
Sz\"{o}ke {\it et al}\/ 1991, Anderson {\it et al}\/ 1992, 
Schafer and Kulander 1992, Potvliege and Smith 1991, 1992a,b, 1994,
Pazdzersky and Yurovsky 1994, Protopapas {\it et al}\/ 1994, 
V\'{e}niard {\it et al}\/ 1995,
Pazdzersky and Usachenko 1997, Fifirig {\it et al}\/ 1997)
and experiment (Muller {\it et al}\/ 1990, Ce Chen and Elliott 1990, 
Baranova {\it et al}\/ 1991, 1992, Yin {\it et al}\/ 1992).
One of the principal points of interest seems to be dependence of the 
observables on the difference of field phases $\varphi$, 
i.e. the problem of {\it phase control}. Another important aspect
is the angular (polar) asymmetry of photoionization rate.
These effects are interrelated since polar asymmetry is
$\varphi$-dependent and vanishes for some particular value
of phase (see more detail below). The calculations have been carried out 
previously for ionization of hydrogen atom in two laser fields 
with a  frequency ratio 1:2 (Schafer and Kulander 1992), 1:3 
(Potvliege and Smith 1991) and 2:3 (Potvliege and Smith 1994).
Potvliege and Smith (1992) presented results for various frequency 
ratios and initial states. Different schemes were employed, but 
all of them implied numerically intensive work.

For the multiphoton electron detachment from negative ions some 
analytical treatment exists (Baranova {\it et al}\/ 1993, 
Pazdzersky and Yurovsky 1994, Pazdzersky and Usachenko 1997) 
being limited mostly to the case when one or both fields 
are weak. The presence of large number of parameters in the problem
sometimes makes results of analytical studies not directly
transparent.
The case of fields with comparable 
(and large) intensities is also of interest bearing in mind both 
possible experimental realizations and the theoretical 
description of the transition between the multiphoton and 
tunneling regimes.

The multiphoton electron detachment from negative ions presents unique
situation when quantitatively reliable results can be obtained
by analytical methods in a broad range of parameters characteristic
to the problem. Indeed, it has been demonstrated recently (Gribakin 
and Kuchiev 1997a,b) that judicious modification of the Keldysh (1964) 
model\footnote{Subsequent development of this model was due to
Perelomov {\it et al}\/ (1966), Faisal (1973) and Reiss (1980);
Perelomov and Popov (1967) were the first to consider multicolour
process within this framework in terms of influence
of higher harmonics on ionization probabilities.}
ensures a very good quantitative agreement with results of 
numerically intensive developments, being much more simple.
In many cases it also provides  good agreement with
available experimental data.
It works unexpectedly well {\it even for small number $n$ of
photons absorbed}. In addition to numerous examples considered
previously (Gribakin and Kuchiev 1997a,b), here we briefly comment
on the most recent experiments by Zhao {\it et al}\/ (1997) on
non-resonant excess photon detachment of negative hydrogen ions.
After absorption of {\it two photons}, the electron is ejected in 
superposition of $S$ and $D$ waves due to selection rules.
The experiment demonstrates prevalence of $D$ wave contribution
(90\% $\pm$ 10\%). Our calculations give for $D$ and $S$ waves 
population 86.2\% and 13.6\% 
respectively\footnote{The model shows also some population of $G$ 
and higher partial waves. However, this unphysical effect proves 
to be less than 0.2\% thus sustaining the model applicability.} 
for experimental conditions (light frequency $\omega = 0.0428$,
field intensity $I=10^{10} {\rm W}/{\rm cm}^2$.
The elaborate numerical calculations by Telnov and Chu (1996)
and by Nikolopoulos and Lambropoulos (1997) give for $D$ wave
population 91\% and 89\% respectively. The difference between
these values is almost the same as the difference between our
result and that of Nikolopoulos and Lambropoulos (1997),
all three theoretical predictions lying within experimental
error bars. This, together with the cases considered earlier
 allows us to suggest that even 
for $n=2$ 
an approach  (Gribakin and Kuchiev 1997a,b)
  provides an  accuracy comparable 
with that of the most elaborate numerical developments. 

The present paper extends approach of Gribakin and Kuchiev (1997a,b)
to the case of bichromatic field. Its objective is to provide 
the scheme and some representative quantitative results for 
two-colour electron detachment from negative ions. In particular,
the phase effects and the polar asymmetry are studied.  
The number of parameters in the problem is quite large and at 
the moment they cannot be  fixed experimentally.
 Nevertheless it seems to be worthwhile
to carry out some pivoting calculations in order to obtain insight
into the possible magnitude of effects specific for negative ions 
in bichromatic fields. We consider angular differential detachment 
rates, heights of ATD (Above Threshold Detachment) peaks 
and total detachment rates.

\section{Scheme of calculations}

The previously developed scheme (Gribakin and Kuchiev 1997a,b)
needs some modifications to incorporate bichromatic problem
when electric field in the light wave
\begin{eqnarray} \label{F}
\vec{F}(t) = \vec{F}_1 \, \cos \omega_1 t +
\vec{F}_2 \, \cos ( \omega_2 t + \varphi )  
\end{eqnarray}
is a superposition of two harmonic components with the frequencies
$\omega_1$, $\omega_2$ and the amplitude vectors $\vec{F}_1$,
$\vec{F}_2$ respectively; $\varphi$ is the difference of field
phases. Atomic units are used throughout the paper unless
stated otherwise.
 
We consider a case of commensurable field 
frequencies\footnote{The general treatment of incommensurable 
frequencies case was considered by Ho {\it et al}\/ (1983), 
Delone {\it et al}\/ (1984), Ho and Chu (1984), 
Manakov {\it et al}\/ (1986), Potvliege and Smith (1992).}
which implies that the common period $T$ exists such that
\begin{eqnarray}
T = \frac{ 2 \pi}{\omega_1} \, M_1 = \frac{ 2 \pi}{\omega_2} \, M_2 
\end{eqnarray}
for some mutually simple integers $M_1$ and $M_2$.

The exact expression for the differential transition rate $d w_\lambda$
is derived following Appendix A of the paper by Gribakin and 
Kuchiev (1997a) with the result
\begin{eqnarray} \label{wd}
d w_\lambda = 2 \pi \sum_{\epsilon_f} 
\left| A_{ \lambda \epsilon_f} \right|^2  
\delta(E_\lambda - E_0 - \epsilon_f) \, \rho_\lambda , 
\\ \label{Ad}
A_{ \lambda \epsilon_f} = \frac{1}{T} \int_0^T \,
\langle \psi_\lambda (t) | V(t) | \psi_0(t) \rangle \, dt .
\end{eqnarray}
Here $\psi_0(t) = \exp(-i E_0 t) \phi_0$ describes an initial state
for the time-independent Hamiltonian $H_0$, and $\psi_\lambda(t)$
is a quasienergy state for the total Hamiltonian $H = H_0 + V(t)$, 
which includes interaction with the periodic field 
$V(t) = - e \vec{r}\cdot \vec{F}(t)$, $V(t) = V(t+T)$:
\begin{eqnarray}
i \frac{ \partial \psi_\lambda}{\partial t} =
\left[ H_0 + V(t) \right] \, \psi_\lambda , \\
\psi_\lambda(t) = \exp( - i E_\lambda t) \, \phi_\lambda ,
\quad \, \quad \quad \phi_\lambda(t) = \phi_\lambda(t+T),
\end{eqnarray}
$E_\lambda$ is the quasienergy, $\rho_\lambda$ is the density of
states, $\vec{r}$ is the active electron vector. The energy
$\epsilon_f$ absorbed by electromagnetic field could be presented
as $\epsilon_f = n_1 \, \omega_1 + n_2 \, \omega_2$ with some integers
$n_1$ and $n_2$. However this representation (i.e. the choice of $n_1$
and $n_2$) generally is non-unique which reflexes existence of different
coherent interfering paths (with different number of absorbed 
and emitted photons of each frequency) leading to the same 
final state. 

If interaction of light wave with an electron is described in the 
dipole-length form, as presumed above, then a long range 
contribution to the matrix elements is emphasized. Therefore it
is sufficient to employ an asymptotic form of the initial-state
wave function (Gribakin and Kuchiev 1997a):
\begin{eqnarray}
 \phi_0(\vec{r}) \approx A r^{\nu-1} \, \exp(- \kappa r) \, 
Y_{lm}(\hat{\vec{r}}) 
\quad \quad \quad 
( r \gg 1),
\end{eqnarray}
where $E_0 = - \frac{1}{2} \kappa^2$, $\nu = Z/\kappa$, $Z$ 
is the charge of the atomic residual core ($\nu=Z=0$ for a 
negative ion), and $\hat{\vec{r}}$ is the unit vector.
The coefficients $A$ are tabulated for many negative ions
(Radzig and Smirnov 1985).

The amplitude $A_{ \lambda \epsilon_f}$ (\ref{Ad}) is evaluated 
neglecting the influence of atomic potential on the photoelectron 
in the final state. Further on, the integral over time in (\ref{Ad}) 
is calculated within the stationary phase approximation which presumes 
large magnitude of the classical action
\begin{eqnarray} \label{S}
S(t) = \frac{1}{2} \int^t \left( \vec{p} + \vec{k}_{t^\prime} 
\right)^2 dt^\prime - E_0 t , 
\end{eqnarray}
where $\vec{k}_t$ is the classical electron momentum due to the
field
\begin{eqnarray} \label{k}
\vec{k}_{t} = e \int^t \vec{F}(t^\prime) \, d t^\prime .
\end{eqnarray}
The photoelectron translational momentum $\vec{p}$ plays a role
of the quantum number $\lambda$ above; in particular, the 
quasienergy $E_\lambda = E_{\vec{p}}$, 
\begin{eqnarray} 
E_{\vec{p}} = \frac{1}{2} \vec{p}^{\: 2} + \frac{e^2}{4 \omega_1^2}F_1^2
+ \frac{e^2}{4 \omega_2^2}F_2^2 
\end{eqnarray}
includes contribution from the electron quiver energy due to
the field.

The result of calculations of the amplitude (\ref{Ad}) could be 
written as a modification of the formula (25) in the paper 
by Gribakin and Kuchiev (1997a):
\begin{eqnarray} \label{A}
A_{ \vec{p} \epsilon_f} = - \frac{(2 \pi)^2}{T} \, A \, \Gamma(1+\nu/2) 
\, 2^{\nu/2} \, \kappa^\nu \, \sum_\mu Y_{lm}(\hat{\vec{p}}_\mu) \,
\frac{\exp \left( i S_\mu \right)}
{\sqrt{2 \pi (- i S^{\prime \prime}_\mu)^{\nu+1}}} . 
\end{eqnarray}
A corresponding  expression for the detachment rate for the negative 
ion case ($\nu = 0$) reads:
\begin{eqnarray} \label{w}
\frac{d w_{e_f}}{d \Omega_{\vec{p}}} = 
\frac{1}{(2 \pi)^2} \, p \left| A_{ \vec{p} \epsilon_f} \right|^2 =
\frac{(2 \pi)^2}{T^2} \, p \, A^2
\left| \sum_\mu Y_{lm}(\hat{\vec{p}}_\mu) \,
\frac{\exp \left( i S_\mu \right)}
{\sqrt{2 \pi S^{\prime \prime}_\mu}} \right|^2 . 
\end{eqnarray}
Here the subscript $\mu$ indicates that the function is calculated
at the saddle point $t= t_\mu$ which satisfies equation
\begin{eqnarray} \label{sp}
S^\prime(t_\mu) = 0 .
\end{eqnarray}
In the plane of the complex-valued time the saddle points $t_\mu$ lie 
symmetrically with respect to the real axis. Summation in (\ref{A})
includes the points lying in the upper half plane 
(${\rm Im} \, t_\mu > 0$) with $ 0 \leq {\rm Re} \, t_\mu \leq T$; 
$\hat{\vec{p}}_\mu$ is a unit vector in the direction of
$\vec{p} + \vec{k}_\mu$.
The magnitude of the final electron translational momentum $p$ is
governed by the energy conservation
\begin{eqnarray} 
\frac{1}{2} \kappa^2 + E_{\vec{p}} = \epsilon_f ,
\end{eqnarray}
which ensures  that the momentum is real for open ATD channels.

According to (\ref{S}), (\ref{k}), (\ref{F}) we have
\begin{eqnarray} 
S^\prime(t) = \frac{1}{2} \, (\vec{p} + \vec{k}_t)^2 + 
\frac{1}{2} \, \kappa^2 =
\nonumber \\ =
\frac{1}{2} \, \vec{p}^{\: 2} + 
\frac{e^2}{2 \omega_1^2}F_1^2 \, \sin^2 \omega_1 t +
\frac{e^2}{2 \omega_2^2}F_2^2 \, \sin^2(\omega_2 t + \varphi) +
\nonumber \\ +
p \cdot F_1 \, \frac{e}{\omega_1} \, \sin \omega_1 t +
p \cdot F_2 \, \frac{e}{\omega_2} \, \sin (\omega_2 t + \varphi) +
\nonumber \\ +
\vec{F}_1 \cdot \vec{F}_2 \, \frac{e^2}{\omega_1 \omega_2} \,
\sin \omega_1 t \, \sin (\omega_2 t + \varphi) + \frac{1}{2} \, \kappa^2 .
\end{eqnarray}

The frequencies $\omega_1$ and $\omega_2$ are integer multiples
of the basic frequency $\omega = 2 \pi / T$
\begin{eqnarray} 
\omega_1 = M_1 \, \omega, \quad \quad \quad \omega_2 = M_2 \, \omega.
\end{eqnarray}
Assuming for definiteness that $M_2 > M_1$ and introducing
$\zeta = \exp ( i \omega t )$ we notice that the function
\begin{eqnarray} 
{\cal P}(\zeta) = \zeta^{2 M_2} S^\prime(\zeta) 
\end{eqnarray}
is a polynomial of the power $4 M_2$ in $\zeta$. 
This observation is of practical importance since equation
(\ref{sp}) defining the saddle point is equivalent to
\begin{eqnarray} \label{pz}
{\cal P}(\zeta) = 0 .
\end{eqnarray}
The efficient numerical procedures for finding roots of polynomials
are available, and, in particular, one can be confident that 
{\it all}\/ the roots are found.

The practical calculations are conveniently carried out using the 
{\it Mathematica}\/ program (Wolfram 1991). Starting from the expression
for $S^\prime(t)$ one derives $S(t)$ and $S^{\prime \prime}(t)$ 
by analytical integration and differentiation respectively.
The saddle points are found using Eq.(\ref{pz}), and the roots
$t_\mu$ lying in the upper half plane are selected. Finite
summation over $\mu$ in (\ref{A}) or (\ref{w}) completes the calculation.

The roots $t_\mu$ and hence the photoionization amplitude (\ref{A})
and rate (\ref{w}) depend on
the orientation of electron translational momentum $\vec{p}$ with
respect to the field amplitudes $\vec{F}_1$ and $\vec{F}_2$. It
is not difficult to consider the fields of various relative 
orientation and polarization, but for simplicity  we limit our
further calculations to linear polarized fields with 
$\vec{F}_1 \parallel \vec{F}_2$. Then the differential photoionization 
rate depends only on the single angle $\theta$ between the vectors
$\vec{p}$ and $\vec{F}_1 \parallel \vec{F}_2$.

\section{Results}

Our calculations for H$^-$ detachment are carried out 
for the parameters $\kappa = 0.2354$, $A=0.75$ (Radzig and Smirnov 1985).
The frequencies ratio $\omega_1 / \omega_2 = 1 : 2$ 
is considered.
In this case the field (\ref{F}) have zero mean value, but possesses 
polar asymmetry (i.e. asymmetry under inversion of the $z$ axis
directed along $\vec{F}_1 \parallel \vec{F}_2$) which could be
conveniently characterized by the time-average value
(Baranova {\it et al}\/ 1990)
\begin{eqnarray} \label{F3}
\langle F^3 \rangle = \frac{3}{4} F_1^2 \, F_2 \, \cos \varphi .
\end{eqnarray}
Presuming that $F_1, \: F_2 > 0$, 
from this expression one infers, 
for instance, that for the phase $\varphi \in [0, \frac{1}{2} \pi]$ 
the electric field effectively attains larger values in the positive-$z$ 
direction than in the negative-$z$ one. This is illustrated, for
example, by Fig.1 in the paper by Schafer and Kulander (1992),
or by Fig.2 in the paper by Baranova {\it et al}\/ (1993).
Note that our definition of the phase $\varphi$ is the same as in
the papers by Baranova {\it et al}\/ (1990), 
Muller {\it et al}\/ (1990),  Pazdzersky and Yurovsky (1995),
but differs from that chosen by 
Schafer and Kulander (1992) who describe the electric filed as
$\vec{F}(t) = \vec{F}_1 \, \sin \omega_1 t +
\vec{F}_2 \, \sin ( \omega_2 t + \varphi_{{\rm KS}})$.
Namely, the phases are related as 
$\varphi_{{\rm KS}}= \varphi - \frac{1}{2}\pi$.

Although the formula (\ref{F3}) shows that the field
has polar symmetry for $\varphi = \frac{1}{2} \pi$ and
the maximal polar asymmetry for $\varphi = 0$, quite
paradoxically, the differential detachment rate (\ref{w})
possesses polar symmetry for $\varphi = 0$ (i.e. is invariant
under transformation $\theta \Rightarrow \pi - \theta$),
and is asymmetrical for other values of phase
(see discussion by Baranova {\it et al}\/ (1990),
Schafer and Kulander (1992), Pazdzersky and Yurovsky (1995)).

The other features of the phase effects are as follows.

\begin{enumerate}

\item
The system Hamiltonian is a 2$\pi$-periodic function of the parameter
$\varphi$.

\item
The system Hamiltonian is  not changed by simultaneous transformation
$\varphi \Rightarrow - \varphi$, $\theta \Rightarrow \pi - \theta$. 
The same applies to the differential detachment rate (\ref{w}).

\item
The transformation $\varphi \Rightarrow \pi - \varphi$ leaves the
Hamiltonian invariant only if $t$ is replaced by $-t$. 

\end{enumerate}

As stressed by Baranova {\it et al}\/ (1990),
Baranova and Zeldovich (1991), Anderson {\it et al}\/ (1992),
Baranova {\it et al}\/ (1993), the problem is invariant
under the time inversion operation provided the final-state 
electron interaction with the atomic core is neglected. 
This is the case in the present model.
Therefore our differential ionization rates are the same for
$\varphi$ and $(\pi - \varphi)$; hence it is sufficient for us to
consider phases only from the interval 
$\varphi \in [0, \: \frac{1}{2} \pi]$.
The calculations by Baranova {\it et al}\/ (1990) within the 
perturbation theory and by Schafer and Kulander (1992) within
the wave propagation technique took into account the
final state electron-core interaction. Therefore they have found some
deviations from the symmetry under $\varphi \Rightarrow (\pi - \varphi)$
transformation. However, for the negative ion photodetachment this
effect could be anticipated to have only minor influence.

At first we consider two fields
with the frequencies $\omega = 0.0043$ and $2\omega$
and equal intensities $I_1 = I_2 = 10^{9} {\rm W}/{\rm cm}^2$.
It is well known that the regime of detachment process is governed 
by the  Keldysh  parameter $\gamma = \omega \kappa / F$
($\gamma \gg1$ for multiphoton detachment in perturbative regime;
$\gamma \ll 1$ for strong field, or tunneling regime). In the present 
case for the first field we have $\gamma_1 = \omega_1 \kappa / F_1 = 6$,
and for the second field $\gamma_2 = 2 \gamma_1$, 
which corresponds to multiphoton regime.

Fig.1 (as well as Figs.2-4 below) shows differential detachment 
rate as a function of the angle $\theta$. The abscissas of the plots 
give the magnitude of
\begin{eqnarray} \label{dd}
\frac{1}{2} \, \frac{1}{2 \pi} \frac{d w_{e_f}}{d \cos \theta} =
\frac{1}{(2 \pi)^2} \, p \left| A_{ \vec{p} \epsilon_f} \right|^2 ,
%\nonumber
\end{eqnarray}
where the right hand side was calculated using the right hand side of 
the formula (\ref{w}). The left hand side of Eq.(\ref{dd}) has the 
factor $\frac{1}{2}$.
It means that the true detachment rate in case of H$^-$ ion is twice
larger than that given by Eq.(\ref{w}).
By this we account for the two possible spin states 
of residual H atom (i.e. for the presence of two equivalent 
electrons in H$^-$). 

In Fig.1 we show the differential detachment rate for the first and 
second ATD peaks which correspond to absorption of 
$n=7$ and $n=8$ photons of frequency $\omega$ respectively. 
In Fig.2 the same results are shown but for doubled
value of the field amplitude $F_2$. In Fig.3 the amplitude
$F_2$ is the same as in Fig.1, but the amplitude $F_1$ is
doubled.

In all cases the angular distributions exhibit
strong dependence on the field phase difference $\varphi$.
This is well expected since the angular patterns
are formed by interference of contributions coming from
different complex-valued moments of time $t_\mu$. 
For $\varphi = 0$ the distribution is rather flat, with $\varphi$
increasing it becomes more oscillatory. An interesting, and not 
obvious feature is that for $\varphi = \frac{1}{2} \pi$
the rate turns zero at the values of angle $\theta$ where it has minimum. 

In Fig.4 we present the results for the same frequencies as before
and equal field intensities $I_1 = I_2 = 10^{11} {\rm W}/{\rm cm}^2$.
Here the Keldysh parameter for the first field is 
$\gamma_1 = \omega_1 \kappa / F_1 = 0.6$.
Bearing in mind  the presence of the second field one can
suppose
that the situation corresponds to the onset of strong-field domain. 
The first open ATD channel corresponds to absorption of $n=18$
photons of frequency $\omega$. The patterns in differential rates 
become more oscillatory than in the weak filed case.

The partial detachment rate for each ATD channel is obtained by 
integration of (\ref{w}) over angles
\begin{eqnarray} \label{wt}
w_{e_f} = \int \frac{d w_{e_f}}{d \Omega_{\vec{p}}} \, 
d \Omega_{\vec{p}} 
= \int_{\theta=\pi}^{\theta=0} 
\frac{d w_{e_f}}{d \cos \theta} \, d \cos \theta 
\: .
\end{eqnarray}
We present separately the result $w_{e_f}^{(u)}$ of integration 
over the upper half-space of the electron ejection 
$\left( 0 < \theta < \frac{1}{2} \pi \right)$ and its counterpart  
$w_{e_f}^{(l)}$ for the lower half-space 
$\left( \frac{1}{2} \pi < \theta < \pi \right)$. These magnitudes  
give a bulk characterization for the partial rate polar asymmetry. 
As discussed above, the polar asymmetry disappears 
(i.e. $w_{N}^{(u)} = w_{N}^{(l)}$) for $\varphi = 0$. 
In the perturbative regime (Fig.5) for the same conditions as 
in Fig.1 we see that the bulk polar asymmetry parameter
$w_{N}^{(u)}/w_{N}^{(l)}$
exceeds $10^3$
even for the lowest ATD channel ($N=1$) provided the phase 
$\varphi$ is not too small (the open ATD channels are labeled by the 
number $N = 1, 2, \ldots$ in order of increasing emitted electron energy). 
For higher ATD peaks the bulk asymmetry is even larger.  

The partial detachment rates integrated over all ejection angles 
$w_N = w_{N}^{(u)} + w_{N}^{(l)}$ is shown in Fig.6
for three representative values of $\varphi$. 
Even for $N=1$ the variation of the phase $\varphi$ leads to 
the substantial change in the detachment 
rate described by a factor  4.

In the tunneling regime the bulk polar asymmetry (Fig.7) is not
as prominent as in the perturbative regime. Nevertheless it is
quite substantial. Note that both in Fig.5 and Fig.7 the electron
emission in the upper half-space 
$\left(0 < \theta < \frac{1}{2} \pi \right)$
is more probable for all $N$, i.e. $w_{N}^{(u)} > w_{N}^{(l)}$  
in the interval of phases considered
($\varphi \in \left[0, \, \frac{1}{2} \pi \right]$).
As discussed above (see the property (ii)), the situation is reversed for 
$\varphi \in \left[- \frac{1}{2} \pi, \, 0 \right]$.
For small value of the phase $\varphi = \frac{1}{8} \pi$ in Fig.7 
there is a clear tendency to swap the relation between 
$w_{N}^{(u)}$ and $w_{N}^{(l)}$ for higher values of $N$ 
which is prevented by a kind of 'pseudocrossing'. For larger
$\varphi$ the partial rates $w_{N}^{(l)}$ are strongly suppressed
when $N$ increases. The phase effects in the partial rates $w_N$
are less significant in the tunneling regime (Fig.8).

The total detachment rates are obtained by summation over
all open ATD channels:
\begin{eqnarray}
w = \sum_{N} w_{N} .
\end{eqnarray}
The results for the total rates as well as 
$w^{(l, \: u)} = \sum_{N} w_{N}^{(l, \: u)}$
are presented in table 1.
In the perturbative regime we see again strong bulk asymmetry
(three orders of magnitude and more) if the phase difference
$\varphi$ is not close to zero, and substantial variation of $w$
with $\varphi$. Actually this result reiterates that
for the partial rate with $N=1$ since the latter gives a
dominant contribution to the total rate in the perturbative regime.

In the strong field regime the bulk polar asymmetry 
$w^{(l)}/w^{(u)}$ remains well manifested in the rate 
summed over all ATD channels.
However, the total rate $w$ is practically insensitive to the phase
variation. The partial rates $w_N$ in Fig.8 exhibit some oscillatory
structure as functions of $N$ with position of extrema depending on
$\varphi$. This $\varphi$-dependence almost completely 
disappears after summation over $N$ as table 1 shows.

\section{Conclusion} 

As a summary, the approach of Gribakin and Kuchiev (1997a,b)
provides convenient 
tool for investigating two-colour photodetachment of negative ions.
The bichromatic electron detachment for H$^-$ ion in the fields
with 1:2 frequency ratio is examined in the perturbative
and tunneling regimes.
The polar asymmetry is found to be
tremendously strong ($  \sim 10^3$) in the perturbative regime.
Note that the asymmetry remains strong and keeps the same
sign for all ATD  for a wide range of phases $0 < \varphi < \pi$.
This property makes this effect be convenient for
experimental observation because 
it manifests itself very strongly  even for a 
relatively poor resolution 
of the phase $\delta \varphi \sim \pi/2$ .
It should be noted that via the recoil mechanism the predicted
effect leads also to
acceleration of the core thus creating  anisotropic flux 
of neutral H atoms.

\section*{Acknowledgments}

We appreciate fruitful discussions with G.F.Gribakin.
One of us (M.Yu.K.) is thankful to the Australian Research Council 
for support. 
This work was supported by the Australian Bilateral Science and 
Technology Collaboration Program.
V.N.O. acknowledges a hospitality of the staff of the 
School of Physics of UNSW where this work was fulfilled.

\section*{References}
\begin{harvard}

\item[] 
Anderson D Z, Baranova N B, Green K, and Zel'dovich B Ya
1992 {\it Zh. Eksp. Teor. Fiz.}\/ {\bf 102} 397-405
[1992 {\it Sov. Phys.-JETP}\/ {\bf 75} 210-4]

\item[] 
Baranova N B, Beterov I M, Zel'dovich B Ya, Ryabtsev I I, 
Chudinov A N and Shul'ginov A A 
1992 {\it Pis'ma Zh. Eksp. Teor. Fiz.}\/ {\bf 55} 431-5
[1992 {\it JETP Letters}\/ {\bf 55} 439-44]

\item[] 
Baranova N B, Zel'dovich B Ya, Chudinov A N and Shul'ginov A A  
1990 {\it Zh. Eksp. Teor. Fiz.}\/ {\bf 98} 1857-68
[1990 {\it Sov. Phys.-JETP}\/ {\bf 71} 1043-9]

\item[] 
Baranova N B and Zel'dovich B Ya 1991 {\it J. Opt. Soc. Am. B}\/
{\bf 8} 27-32

\item[] 
Baranova N B, Reiss H R and Zel'dovich B Ya 1993 {\it Phys. Rev. A}\/
{\bf 48} 1497-505 

\item[]
Ce Chen and Elliott D S 1990 {\it Phys. Rev. Lett.}\/ {\bf 65} 1737-40 

\item[]
Delone N B, Manakov N L and Fainshtein A G 1984
{\it Zh. Eksp. Teor. Fiz.}\/ {\bf 86} 906-14
[1984 {\it Sov. Phys.-JETP}\/ {\bf 59} 529-33]

\item[] 
Gribakin G F and Kuchiev M Yu 1997a {\it Phys. Rev. A}\/
{\bf 55} 3760-71 

\item[] 
\dash 1997b {\it J. Phys. B: At. Mol. Opt. Phys.}\/ {\bf 30} L657-64

\item[] 
Faisal F H M 1973 {\it J. Phys. B: At. Mol. Phys.}\/ {\bf 6} L89-92

\item[] 
Fifirig M, Cionga A and Florescu V 1997 {\it J. Phys. B: At. Mol. 
Opt. Phys.}\/ {\bf 30} 2599-608

\item[]
Ho T S, Chu S I and Tietz J V 1983 {\it Chem. Phys. Lett.}\/
{\bf 96} 464-71

\item[]
Ho T S and Chu S I 1984 {\it J. Phys. B: At. Mol. Opt. Phys.}\/ 
{\bf 17} 2101-28

\item[]
Keldysh L V 1964 {\it Zh. Eksp. Teor. Fiz.}\/ {\bf 47} 1945-57
[1965 {\it Sov. Phys.-JETP}\/ {\bf 20} 1307-14]

\item[]
Nikolopoulos L A A and Lambropoulos P 1997 {\it Phys. Rev. A}\/ 
{\bf 56} 3106-15

\item[]
Manakov N L, Ovsiannikov V D and Rapoport L P 1986 {\it Phys.Rep.}
{\bf 141} 319-433

\item[] 
Muller H G, Bucksbaum P H, Schumacher D W and Zavriev A
1990 {\it J. Phys. B: At. Mol. Opt. Phys.}\/ {\bf 23} 2761-9

\item[] 
Pazdzersky V A and Yurovsky V A 1991
{\it J. Phys. B: At. Mol. Opt. Phys.}\/ {\bf 24} 733-40

\item[] 
\dash 1994 {\it Phys. Rev. A}\/ {\bf 51} 632-40

\item[] 
Pazdzersky V A and Usachenko V I 1997 {\it J. Phys. B: At. Mol. Opt. Phys.}\/
{\bf 30} 3387-402

\item[]
Perelomov A M, Popov V S and Terent'ev M V 1966 {\it Zh. Eksp. Teor. Fiz.}\/ 
{\bf 50} 1393-409 [1966 {\it Sov. Phys.-JETP}\/ {\bf 23} 924-34]

\item[]
Perelomov A M and Popov V S 1967 {\it Zh. Eksp. Teor. Fiz.}\/ 
{\bf 52} 514-26 [1967 {\it Sov. Phys.-JETP}\/ {\bf 25} 336-43]

\item[] 
Potvliege R M and Smith  P H G 1991 {\it J. Phys. B: At. Mol. Opt. Phys.}\/
{\bf 24} L641-6

\item[] 
\dash 1992 {\it J. Phys. B: At. Mol. Opt. Phys.}\/ {\bf 25} 2501-16

\item[] 
\dash 1994 {\it Phys. Rev. A}\/ {\bf 49} 3110-3

\item[] 
Protopapas M, Knight P L and Burnett K 1994 {\it Phys. Rev. A}\/ 
{\bf 49} 1945-9

\item[]
Radzig A A and Smirnov B M 1985 {\it Reference Data on Atoms,
Molecules and Ions} (Berlin: Springer)

\item[] 
Reiss H R 1980 {\it Phys. Rev. A}\/ {\bf 22} 1786-813

\item[] 
Schafer K J and Kulander K C 1992 {\it Phys. Rev. A}\/
{\bf 45} 8026-33 

\item[]
Sz\"{o}ke A, Kulander K C and Bardsley J N 1991 {\it J. Phys. B: 
At. Mol. Opt. Phys.}\/ {\bf 24} 3165-71

\item[]
Telnov D A and Chu S I 1996 {\it J. Phys. B: 
At. Mol. Opt. Phys.}\/ {\bf 29} 4401-10

\item[]
V\'{e}niard V, Taleb R and Maquet A 1995 {\it Phys. Rev. Lett.}\/
{\bf 74} 4161-4 

\item[]
Wolfram S 1991 {\it Mathematica: A System for Doing Mathematics by
Computer}, 2nd ed. (Addison-Wesley Publishing Co., Palo Alto)

\item[] 
Yin Y-Y, Ce Chen and Elliott D S 1992 {\it Phys. Rev. Lett.}\/ {\bf 69}
2353-6

\item[] 
Zhao X M, Gulley M S, Bryant H C, Strauss C E M, Funk D J, Stintz A,
Rislove D C, Kyrala G A, Ingalls W B and Miller W A 1997
{\it Phys. Rev. Lett.}\/ {\bf 78} 1656-9

\end{harvard}

\newpage

\begin{table} 
\caption{
Total rates $w$ (summed over all ATD channels) for detachment of H$^-$
ion in the bichromatic field with the frequencies 
$\omega = 0.0043$ and $2\omega$, equal intensities 
$I_1 = I_2$ and some representative values of the phase
difference $\varphi$. The detachment rates $w^{(u)}$ and $w^{(l)}$ 
for electron ejection into the upper and lower half-spaces
respectively are also shown.}

\vspace{5mm}
\begin{tabular}{| c || l | l | l || l | r | l |}
\hline
& & & & & & \\
$\varphi$ & $w$ & $w^{(u)}$ & $w^{(l)}$ & 
$w$ & $w^{(u)}$ & $w^{(l)}$ \\
& & & & & & \\
\hline
& \multicolumn{3}{c ||}{} & \multicolumn{3}{c |}{} \\
& \multicolumn{3}{c ||}{in units 10$^{-21}\:$a.u.} & 
\multicolumn{3}{c |} {in units 10$^{-6}\:$a.u.} \\
& \multicolumn{3}{c ||}{} & \multicolumn{3}{c |}{} \\
\hline
\hline
& \multicolumn{3}{c ||}{} & \multicolumn{3}{c |}{} \\
& \multicolumn{3}{c ||}{$I_1 = I_2 = 10^9$W/cm$^2$} & 
\multicolumn{3}{c |} {$I_1 = I_2 = 10^{11}$W/cm$^2$} \\
& \multicolumn{3}{c ||}{} & \multicolumn{3}{c |}{} \\
\hline
& & & & & & \\
0 & $1.55$ & 0.773 & 0.773 & 162.1 & 81.1 & 81.1 \\
& & & & & & \\
\hline
& & & & & & \\
$\frac{1}{8} \pi$ & 2.12 & 1.88 & 0.240 & 164.6 & 99.8 & 64.8 \\
& & & & & & \\
\hline
& & & & & & \\
$\frac{1}{4} \pi$ & 3.56 & 3.50 & 0.00566 & 164.8 & 113.0 & 51.8 \\
& & & & & & \\
\hline
& & & & & & \\
$\frac{3}{8} \pi$ & 5.07 & 5.06 & 0.00112 & 165.1 & 123.8 & 41.4 \\
& & & & & & \\
\hline
& & & & & & \\
$\frac{1}{2} \pi$ & 5.71 & 5.71 & 0.00040 & 164.9 & 128.0 & 37.0 \\
& & & & & & \\
\hline
\end{tabular}
\end{table}

\Figures

%\begin{figure}
%\caption{
%Detachment of H$^-$ ion in bichromatic field 
%with the frequencies $\omega = 0.0043$ and $2\omega$
%and equal intensities $I_1 = I_2 = 10^{9} {\rm W}/{\rm cm}^2$.
%Differential detachment rate (see Eq.(\ref{dd}))
%(in units $10^{-12}$a.u.) as a function of the angle $\theta$ 
%is shown for the first ({\it a}\/, absorption of $n=7$ photons
%of frequency $\omega$) and second ({\it b}\/, $n=8$) ATD peaks
%and various values of the field phase difference: 
%solid curve - $\varphi = 0$;
%short-dashed curve - $\varphi = \frac{1}{8} \pi$;
%dot-dashed curve - $\varphi = \frac{1}{4} \pi$;
%dotted curve - $\varphi = \frac{3}{8} \pi$;
%long-dashed curve - $\varphi = \frac{1}{2} \pi$.}
%\end{figure}

\begin{figure}
\caption{
Detachment of H$^-$ ion in bichromatic field 
with the frequencies $\omega = 0.0043$ and $2\omega$
and equal intensities $I_1 = I_2 = 10^{9} {\rm W}/{\rm cm}^2$.
Differential detachment rate (see Eq.(\protect\ref{dd}))
(in units $10^{-12}$a.u.) as a function of the angle $\theta$ 
is shown for the first ({\it a}\/, absorption of $n=7$ photons
of frequency $\omega$) and second ({\it b}\/, $n=8$) ATD peaks
and various values of the field phase difference: 
solid curve - $\varphi = 0$;
short-dashed curve - $\varphi = \frac{1}{8} \pi$;
dot-dashed curve - $\varphi = \frac{1}{4} \pi$;
dotted curve - $\varphi = \frac{3}{8} \pi$;
long-dashed curve - $\varphi = \frac{1}{2} \pi$.}
\end{figure}

\begin{figure}
\caption{
Same as in Fig.1, but for unequal field  
intensities $I_1 = 10^{9} {\rm W}/{\rm cm}^2$,
$I_1 = 4 \times 10^{9} {\rm W}/{\rm cm}^2$.
The differential detachment rates is shown in units $10^{-10}$a.u.}
\end{figure}

\begin{figure}
\caption{
Same as in Fig.1, but for unequal field  
intensities $I_1 = 4 \times 10^{9} {\rm W}/{\rm cm}^2$,
$I_1 = 10^{9} {\rm W}/{\rm cm}^2$.
The differential detachment rate is shown in units $10^{-10}$a.u.}
\end{figure}

\begin{figure}
\caption{
Same as in Fig.1, but in the strong field regime:  
$I_1 = I_2 = 10^{11} {\rm W}/{\rm cm}^2$.
The detachment rate for the first ({\it a}\/, absorption of 
$n=18$ photons of frequency $\omega$) and second 
({\it b}\/, $n=19$) ATD peaks is shown in units $10^{-6}$a.u.}
\end{figure}

\begin{figure}
\caption{
Partial detachment rates for various ATD channels for H$^-$ ion in 
bichromatic field with the same parameters as in Fig.1
(perturbative regime) and various values of the field 
phase difference $\varphi$.
$N$ labels ATD peaks with the lowest $N=1$ peak corresponding
to absorption of 7 photons of frequency $\omega = 0.0043$.
The rate $w_{N}^{(u)}$ of electron emission in the upper
half-space is shown by circles, its counterpart 
$w_{N}^{(l)}$ for the lower half-space is depicted by triangles.
The plot for $\varphi = \frac{1}{8} \pi$ additionally
includes the rate for $\varphi = 0$ (crosses) when 
emission is polar symmetrical ($w_{N}^{(u)} = w_{N}^{(l)}$).
The symbols are joined by lines to help the eye.}
\end{figure}

\begin{figure}
\caption{
Same as in Fig.5, but for the detachment rate integrated over all 
angles $w_N = w_{N}^{(u)} + w_{N}^{(l)}$. 
The results for three values of the phase $\varphi$
are shown: circles -- $\varphi = 0$;
blocks -- $\varphi = \frac{1}{4} \pi$, 
triangles -- $\varphi = \frac{1}{2} \pi$.}
\end{figure}

\begin{figure}
\caption{
Same as in Fig.5, but for the field parameters chosen as in Fig.4
(strong field regime). The lowest $N=1$ peak corresponds
to absorption of 18 photons of frequency $\omega = 0.0043$.
The detachment rate is shown in units $10^{-6}$ a.u.}
\end{figure}

\begin{figure}
\caption{
Same as in Fig.6, but for the field parameters chosen as in Fig.7
(strong field regime). The detachment rate is shown in units 
$10^{-6}$ a.u.}
\end{figure}

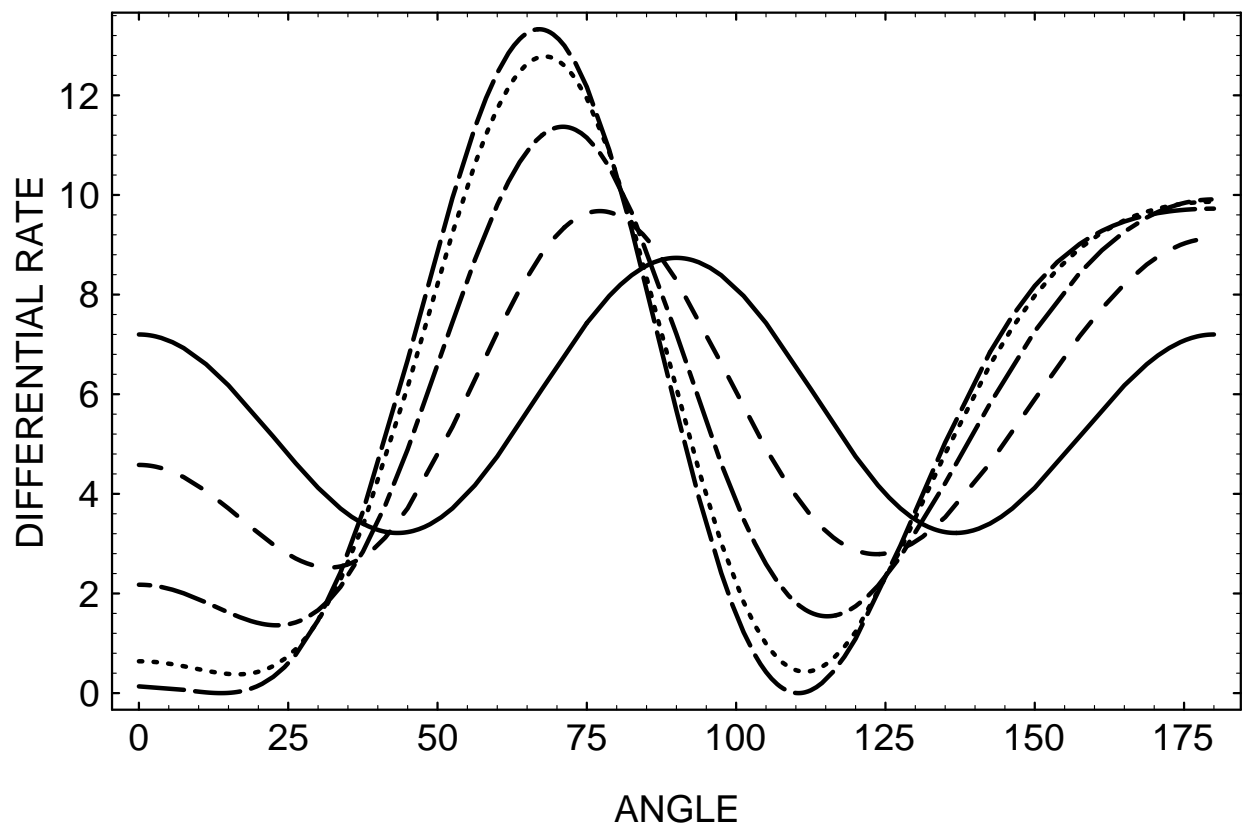
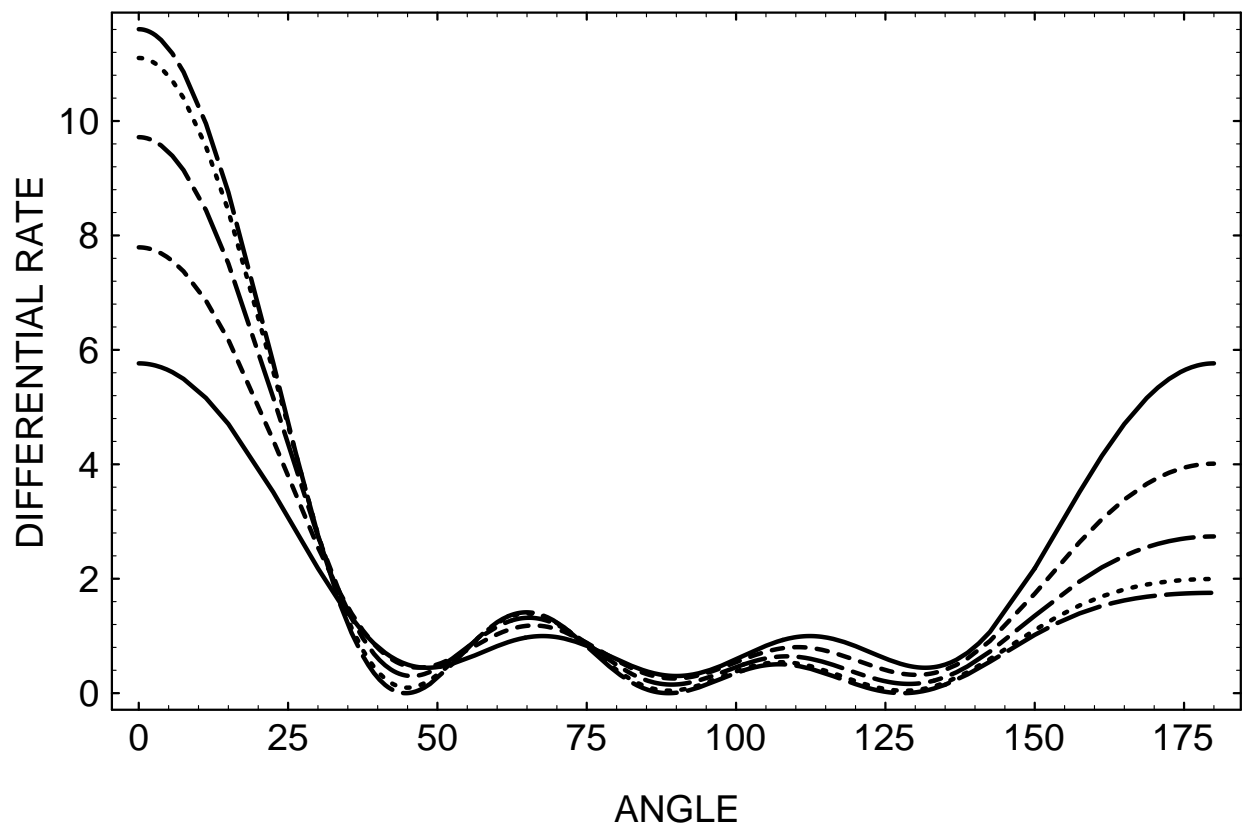
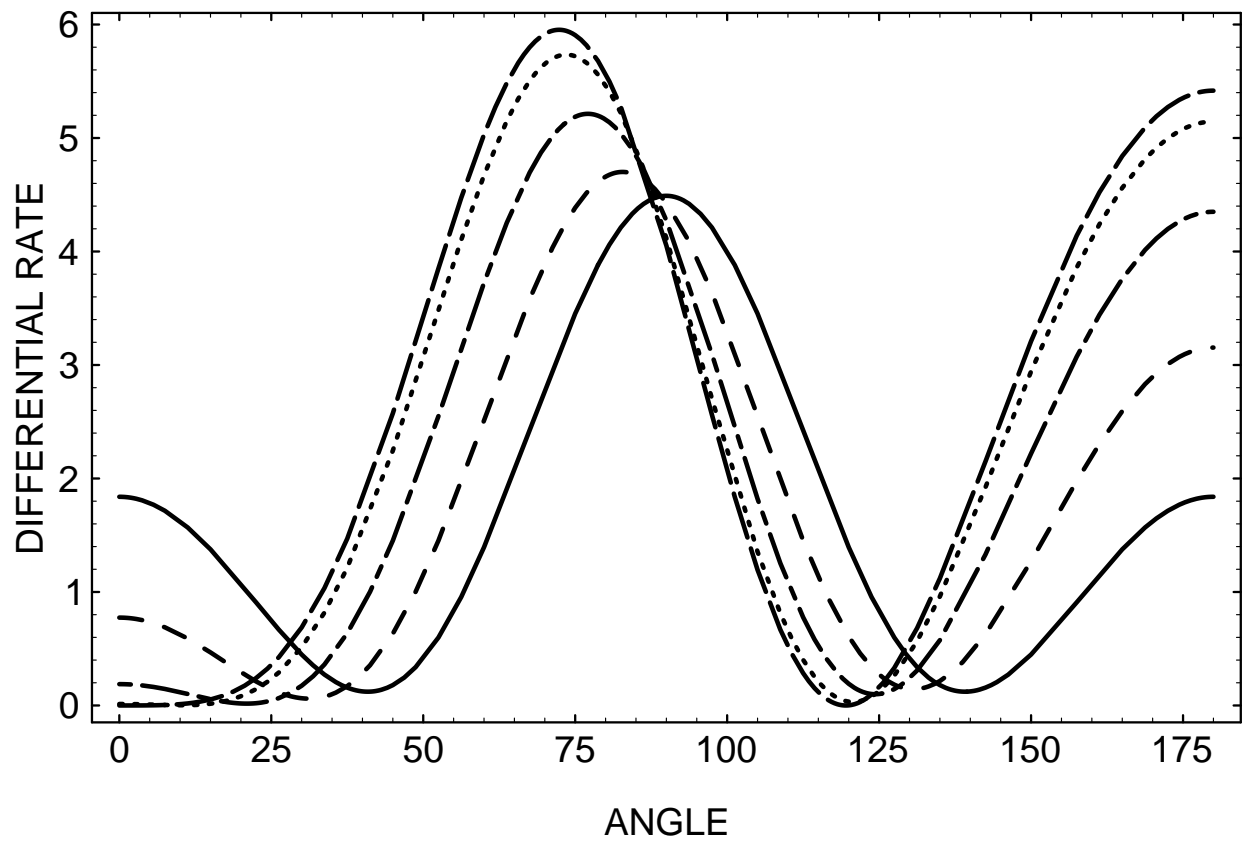
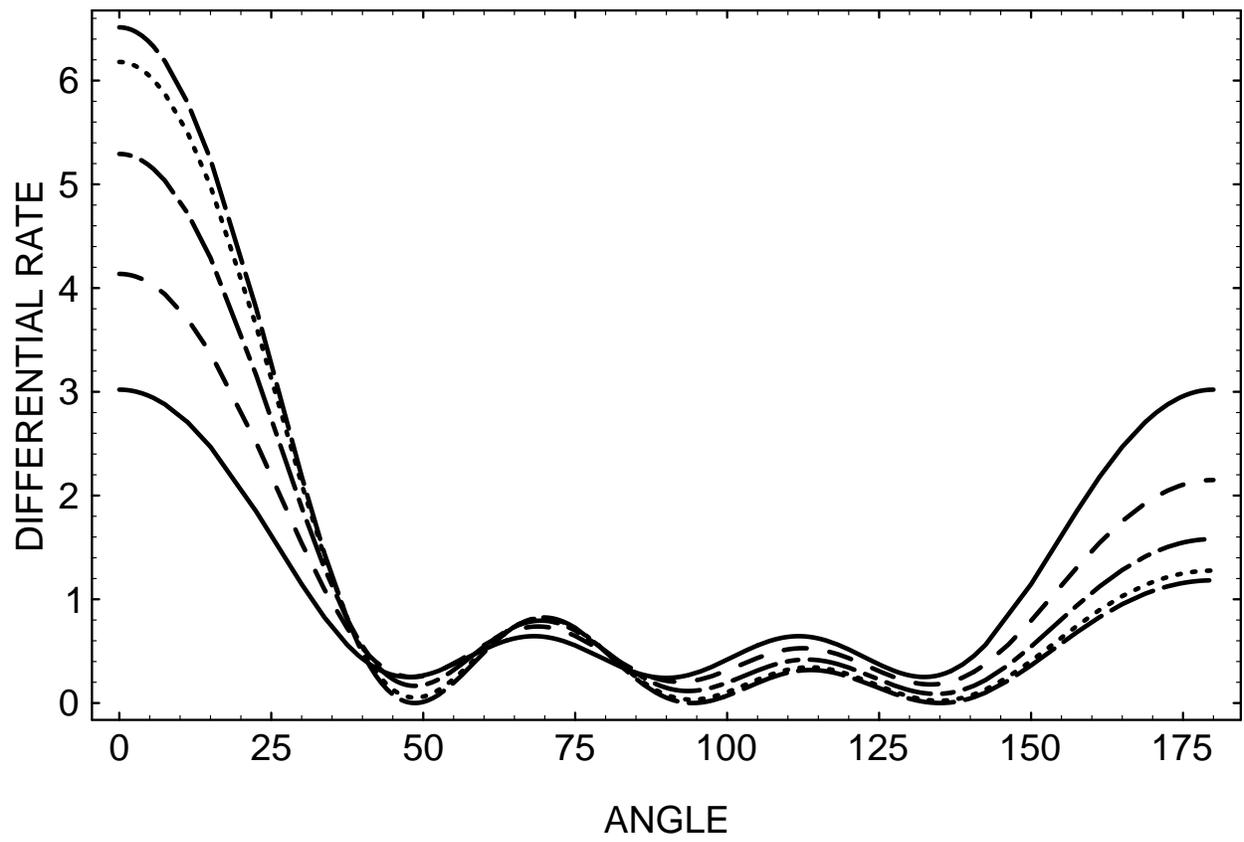
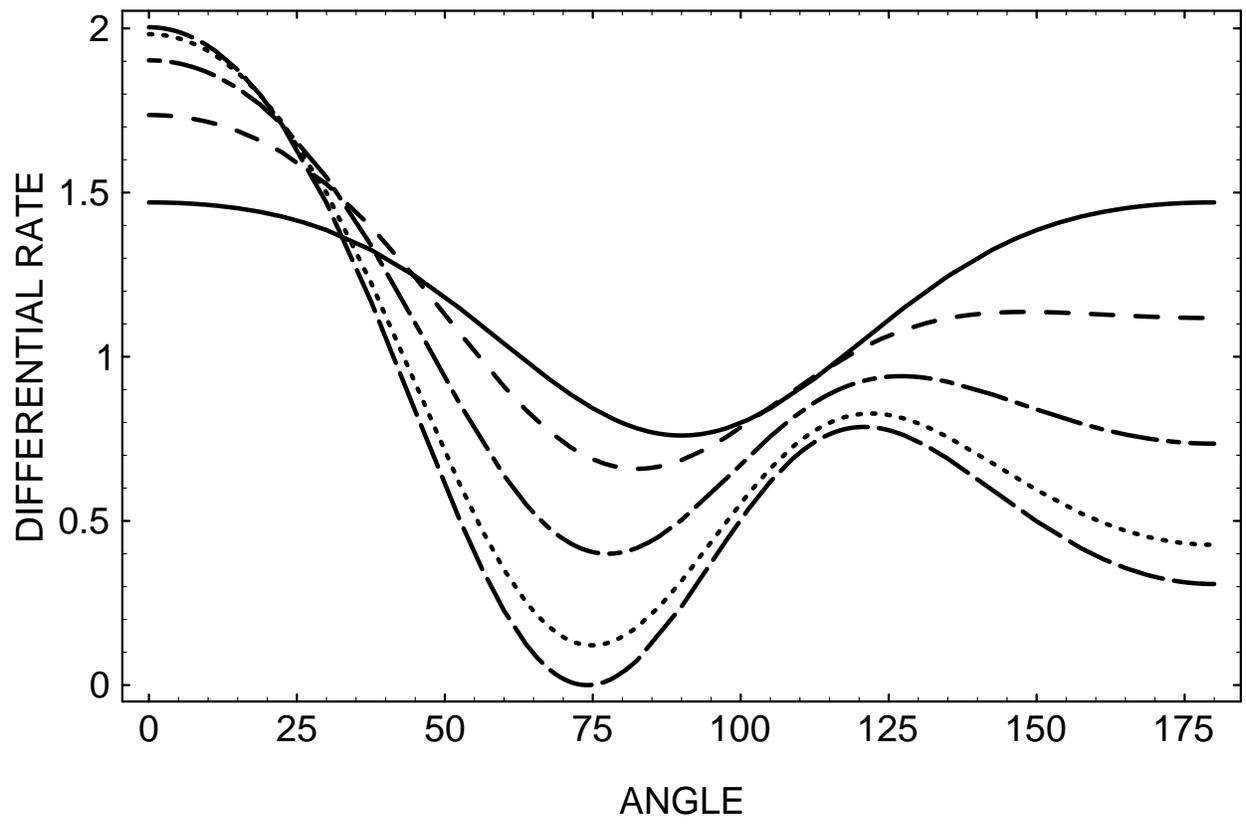
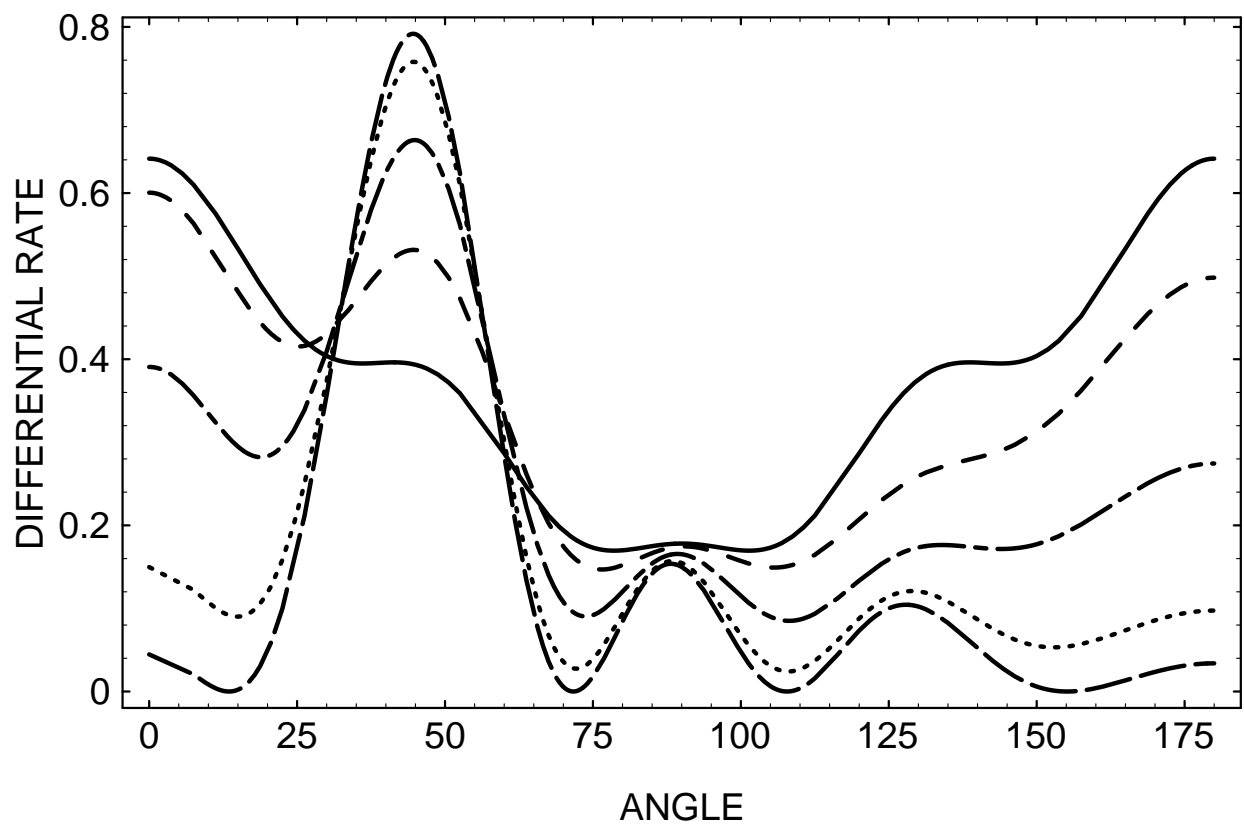
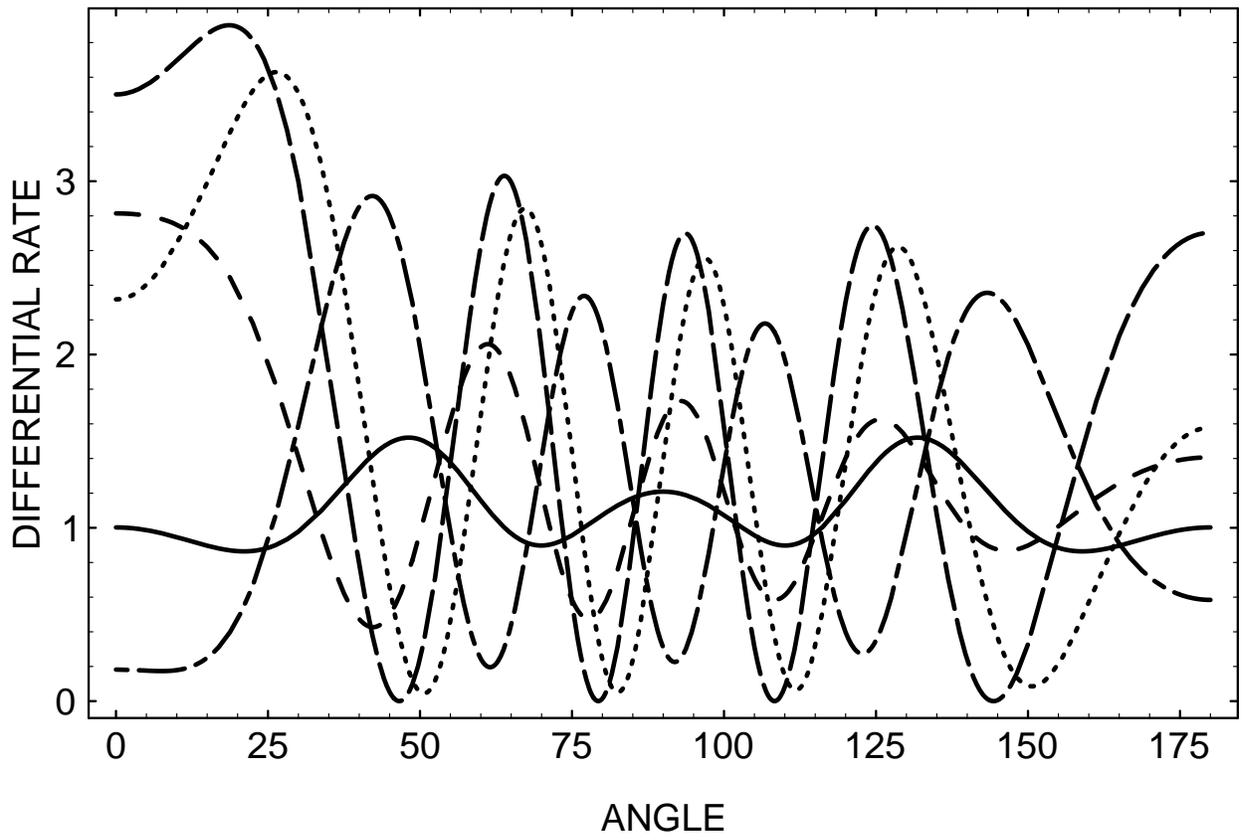
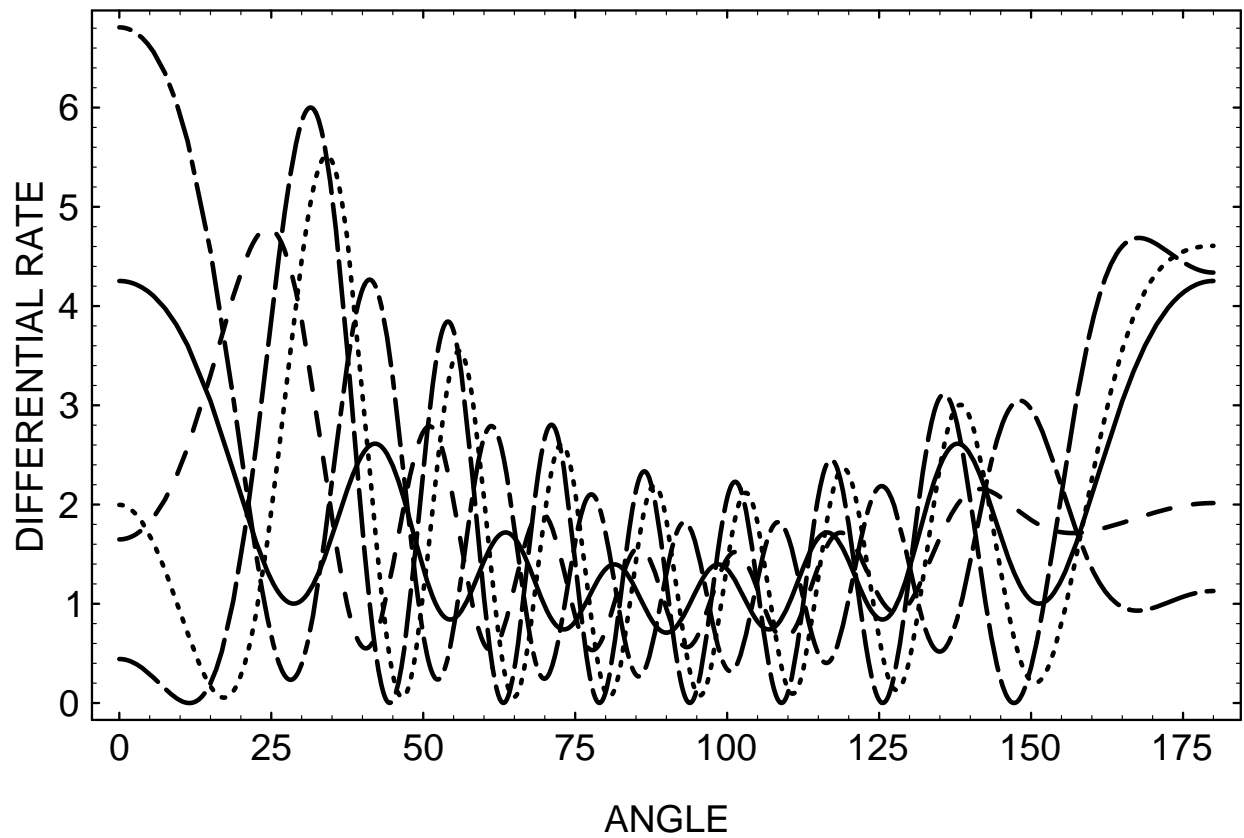
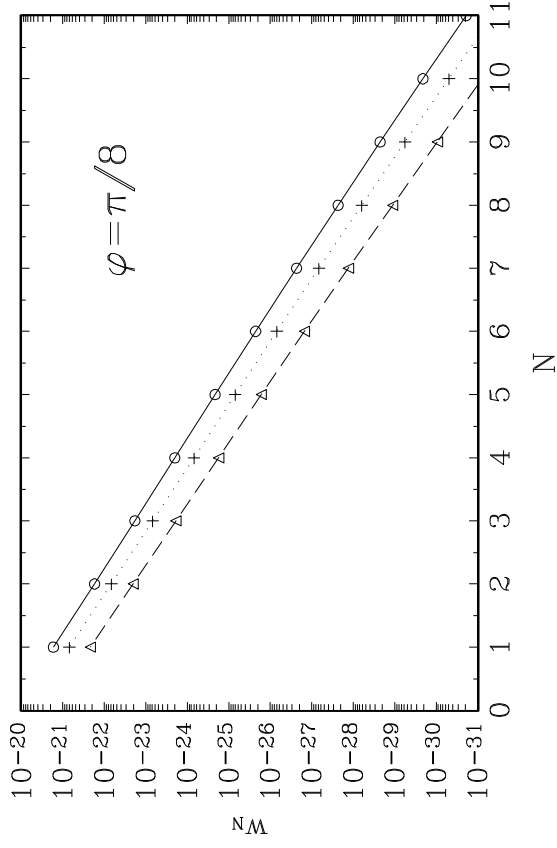
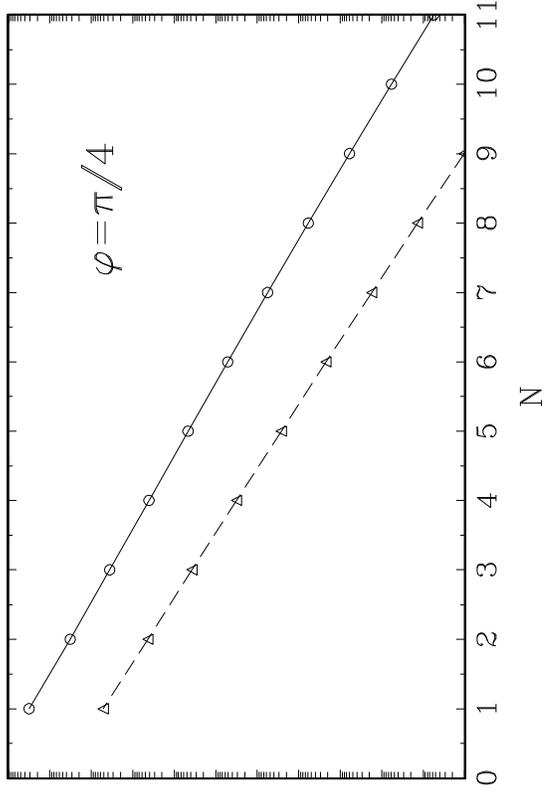
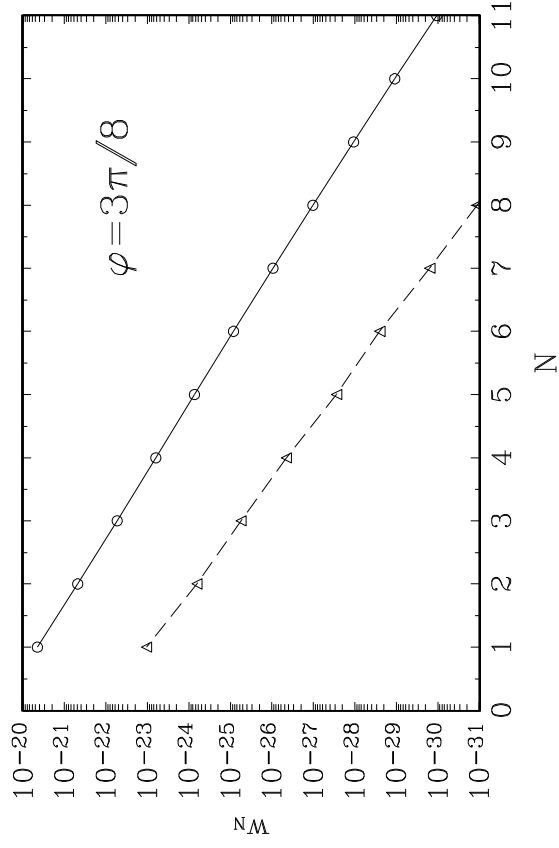
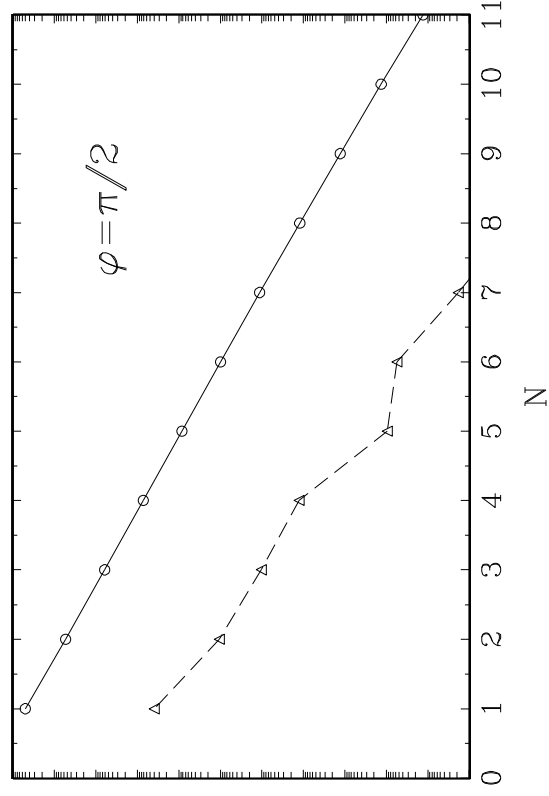
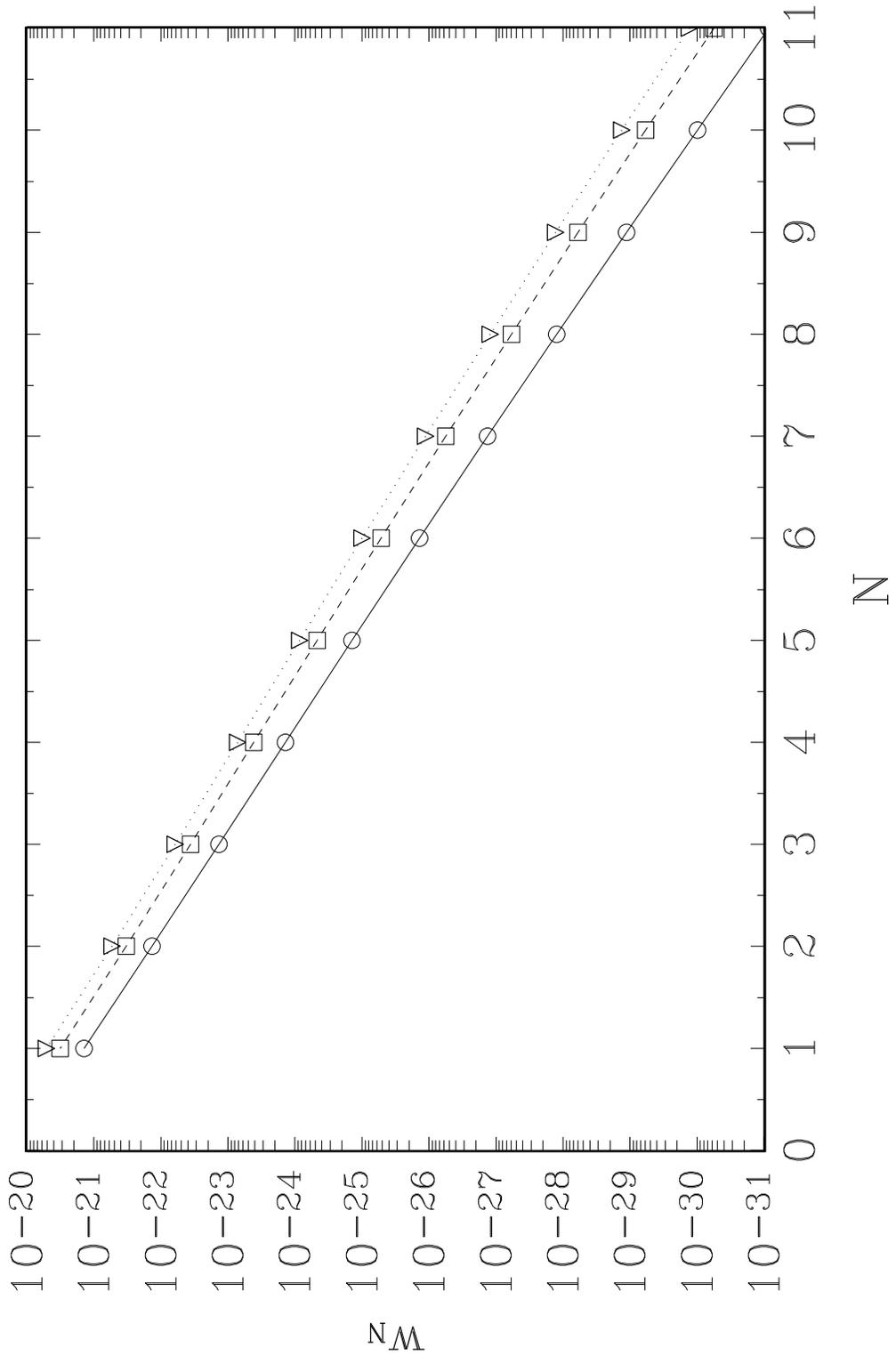
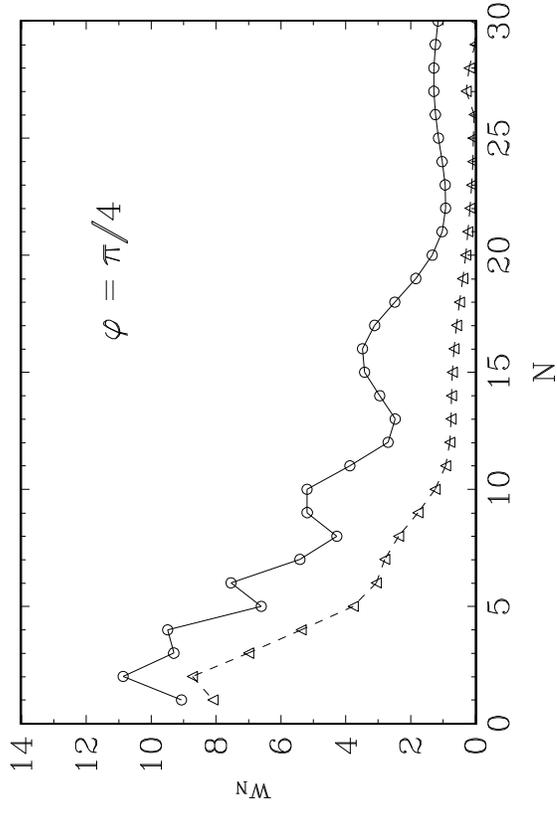
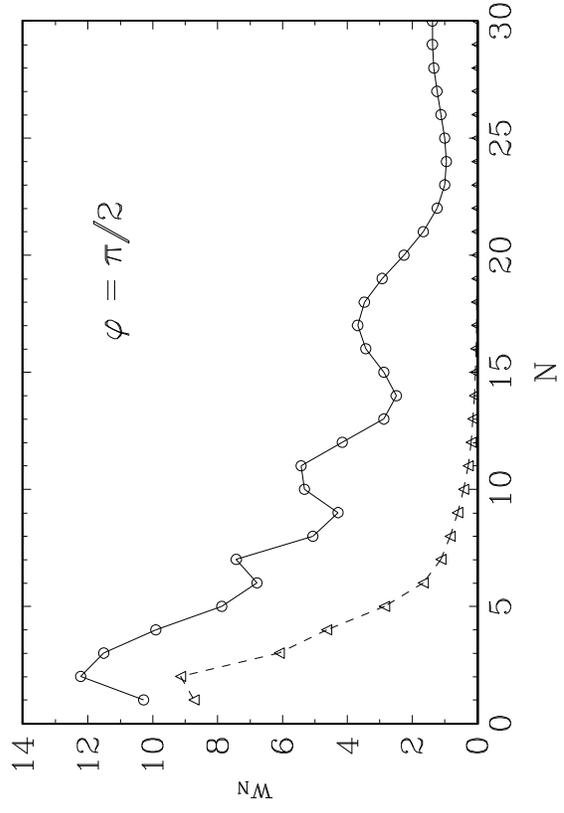
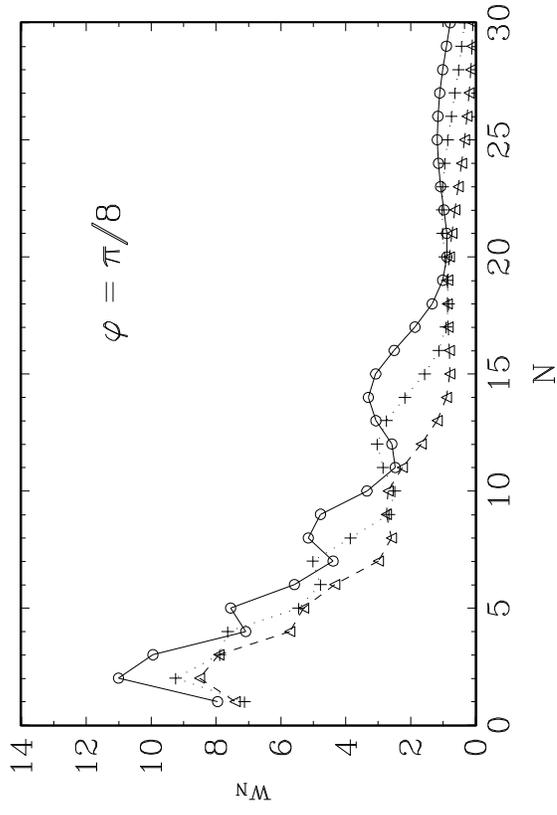
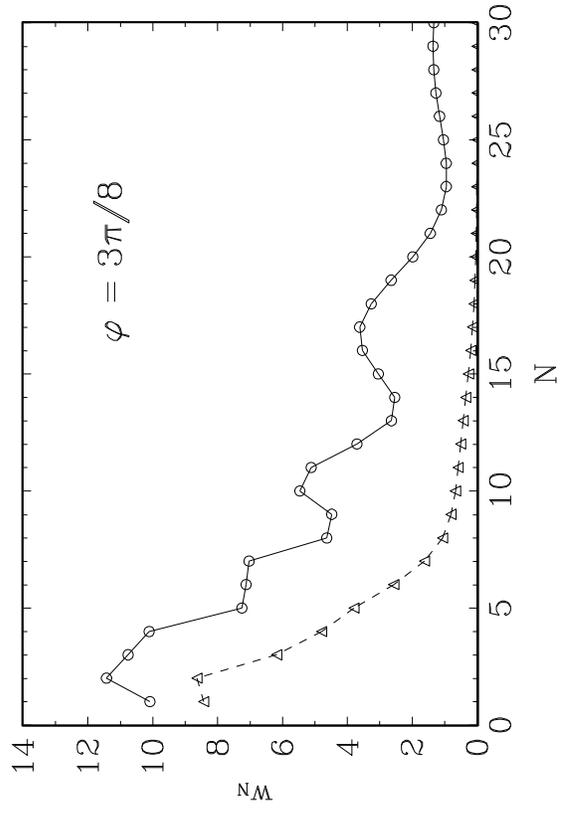
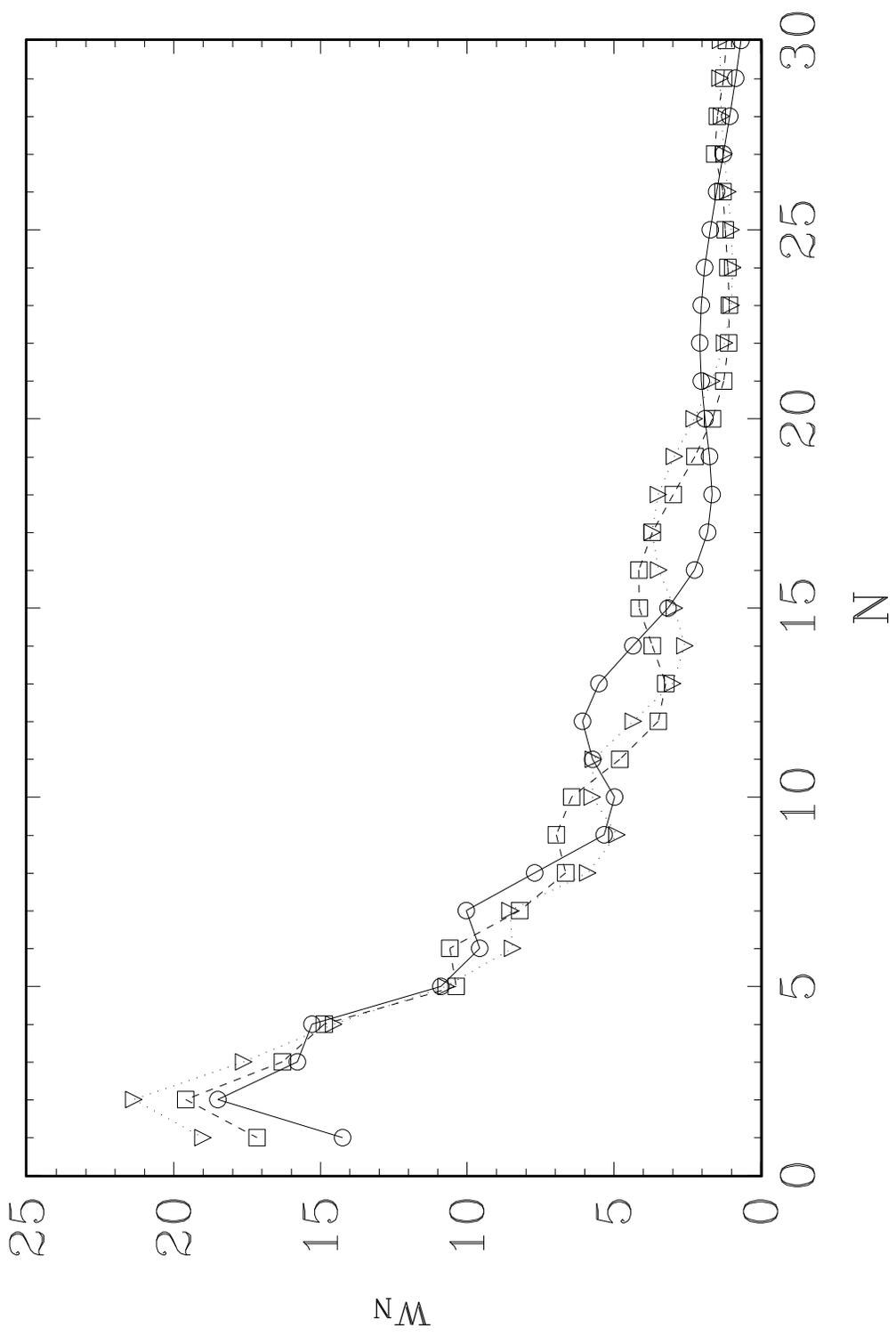
\end{document}